\documentclass[%
 preprint,
 superscriptaddress,
 amsmath,amssymb,
 aps,
 showkeys,
 titlepage,
 endfloats*.   
]{revtex4-2}

\usepackage[utf8]{inputenc}
\usepackage[english]{babel}

\usepackage{graphicx}
\usepackage{dcolumn}
\usepackage{bm}

\usepackage[colorlinks = true,
            linkcolor = blue,
            urlcolor  = blue,
            citecolor = red,
            anchorcolor = blue]{hyperref}

\begin{document}



\title{Electron Drag Effect on Thermal Conductivity in Two-dimensional Semiconductors} 

\author{Yujie Quan}
\affiliation{Department of Mechanical Engineering, University of California, Santa Barbara, CA 93106, USA}

\author{Bolin Liao}
\email{bliao@ucsb.edu} \affiliation{Department of Mechanical Engineering, University of California, Santa Barbara, CA 93106, USA}

\date{\today}

\begin{abstract}
Two-dimensional (2D) materials have shown great potential in applications as transistors, where thermal dissipation becomes crucial because of the increasing energy density. 
Although thermal conductivity of 2D materials has been extensively studied, interactions between nonequilibrium electrons and phonons, which can be strong when high electric fields and heat current coexist, are not considered. 
In this work, we systematically study the electron drag effect, where nonequilibrium electrons impart momenta to phonons and influence the thermal conductivity, in 2D semiconductors using ab initio simulations. 
We find that, at room temperature, electron drag can significantly increase thermal conductivity by decreasing phonon-electron scattering in 2D semiconductors, while its impact in three-dimensional (3D) semiconductors is negligible.
We attribute this difference to the large electron-phonon scattering phase space and higher contribution to thermal conductivity by drag-active phonons.
Our work elucidates the fundamental physics underlying coupled electron-phonon transport in materials of various dimensionalities.

\end{abstract}

                            
\maketitle


\section{Introduction}
Two-dimensional (2D) materials have attracted great interest since the discovery of graphene \cite{novoselov2004electric} due to their excellent mechanical, electrical, and optical properties \cite{akinwande2017review,ma2021tunable,novoselov20162d}. 
Nowadays, they are widely used in electronic \cite{kim2015materials,huang20222d}, optoelectronic \cite{yang20182d,long2019progress}, and energy conservation and storage devices \cite{zhang20162d,glavin2020emerging}. 
Among these, 2D materials-based transistors show tremendous potential in downsizing three-dimensional (3D) transistors, which is unachievable due to the severe decrease of charge mobility caused by the increased scatterings in ultrathin channels \cite{chhowalla2016two,sebastian2021benchmarking,das2021transistors}. 
Unlike 3D materials, the dangling-bond-free surface of 2D materials suppresses the generation of interfacial trap states and thus preserves excellent electrical properties \cite{huang20222d}. 
Moreover, the atomically uniform thickness enables heterogeneous integration and minimizes variability across different devices.
As device miniaturization progresses, the power density increases substantially, making efficient thermal management strategies for long-term stability necessary.
Therefore, a thorough theoretical understanding of thermal transport properties in 2D materials becomes crucial for devising effective thermal dissipation solutions in devices~\cite{gu2018colloquium}.

There have been extensive theoretical studies on the thermal transport properties of 2D materials. 
Thanks to the development in \textit{ab initio} methods, thermal transport properties can be obtained by solving the phonon Boltzmann transport equation (BTE) using interatomic force constants (IFCs) obtained from density functional theory (DFT) and density functional perturbation theory (DFPT). 
Among all 2D materials, monolayer graphene was found to have the highest thermal conductivity \cite{lindsay2014phonon,feng2018four,han2023thermal}. 
The calculated thermal conductivity reaches 3000\,W/mK at room temperature if only the three-phonon scattering is considered \cite{lindsay2014phonon} but this value becomes much smaller if the four-phonon scattering is included \cite{feng2018four,han2023thermal}.
Besides, 2D materials also possess distinct phonon transport properties. 
Because of the dominant momentum conserving normal scattering and large density of states of long-wavelength flexural acoustic (ZA) phonons, hydrodynamic phonon transport is predicted to occur in graphene, boron nitride and other 2D materials at significantly higher temperatures and wider temperature ranges compared with 3D bulk counterparts \cite{lee2015hydrodynamic,cepellotti2015phonon,yang2019hydrodynamic}. This effect was subsequently observed experimentally~\cite{huberman2019observation}. Besides hydrodynamic phonon transport, coherent phonon effect~\cite{choudhry2019origins} and surface-enhanced resonant bonding~\cite{yue2018ultralow} are also observed to strongly impact thermal transport in 2D materials.
In addition to the understanding of pristine crystals, thermal transport properties in doped 2D materials are also widely studied. 
It is found that the thermal conductivity of silicene can be reduced by over 40\% due to phonon-electron scatterings with carrier concentration around $\rm 10^{13}~cm^{-2}$, while phonon-electron scatterings have less impact on the thermal conductivity of phosphorene \cite{yue2019controlling}. 
For materials with symmetry of reflection to the basal plane, such as graphene, ZA phonons cannot directly interact with electrons \cite{fischetti2016mermin}. 
However, it is found that the thermal conductivity of graphene can be significantly reduced due to the indirect interaction between ZA phonons and electrons, mediated by in-plane transverse acoustic (TA) and longitudinal acoustic (LA) phonons \cite{yang2021indirect}.
Despite the remarkable advancement in theoretical understanding of thermal transport in doped 2D materials, the current method, referring to the assumption that electrons are in equilibrium states when calculating phonon distribution functions and vice versa, ignores mutual interactions between out-of-equilibrium electrons and phonons during transport.
Given the promising potential of 2D materials-based transistors, where electric fields and heat currents coexist, investigating the effect of nonequilibrium electrons on phonon transport and thermal conductivity is critical for efficient thermal management design. 

Momentum exchange between nonequilibrium electrons and phonons is generally referred to as the ``electron-phonon drag'' effect \cite{gurevich1989electron,herring1954theory}. 
The phonon drag effect was first observed in the prominent increase of the Seebeck coefficient in Ge \cite{frederikse1953thermoelectric,geballe1954seebeckGe}, Si \cite{geballe1955seebeckSi}, and $\rm FeSb_2$ \cite{bentien2007colossalFeS2,pokharel2013phononFeS2,battiato2015unifiedFeS2} at very low temperatures.
Later, it was found that, at room temperature, the phonon drag contribution to the Seebeck coefficient can be significant in Si by solving the partially decoupled electron BTE \cite{mahan2014seebeck,zhou2015ab,fiorentini2016thermoelectric}.
Further, Protik et al. developed a numerical scheme to efficiently solve the fully coupled electron-phonon BTEs using the inputs from DFT and DFPT calculations to capture the electron-phonon drag effect on transport properties \cite{protik2022elphbolt}.
Using this method, they demonstrated a large phonon drag contribution to the Seebeck coefficient in SiC \cite{protik2020electronSiC}, GaAs \cite{protik2020coupledGaAs}, diamond \cite{li2022colossal}, and BAs \cite{li2023high} at room temperature. 
Along this line, we have recently discovered that, at room temperature, phonon drag not only significantly increases the Seebeck coefficient but also enhances carrier mobility in wide bandgap materials GaN and AlN \cite{quan2023significant}.
While the phonon drag effect has received extensive study, little attention has been given to the electron drag effect, i.e., the influence of nonequilibrium electrons on thermal transport properties mediated by phonons. 
Protik et al. calculated the electron drag effect on the thermal conductivity of 3D GaAs \cite{protik2020coupledGaAs} and SiC \cite{protik2020electronSiC}, but they found this effect is quite minimal, leading to less than 1\% thermal conductivity change. 
However, given the distinct electron and phonon transport properties in 2D materials from their 3D counterparts, it remains unclear whether the electron drag effect in 2D will differ appreciably compared to that in 3D bulk materials.

In this work, we use the recently developed fully coupled electron-phonon BTE solver \cite{protik2022elphbolt} to investigate the impact of electron drag on thermal transport properties in 2D semiconductors from first principles. For a systematic comparison, we pick three representative material systems for this study: MoS$_2$, which is a polar 2D semiconductor without inversion symmetry; phosphorene, which is a nonpolar 2D semiconductor with inversion symmetry; and AlN, which is a polar 3D semiconductor without inversion symmetry. Inversion symmetry is important here because of the strong piezoelectric scattering of long-wavelength acoustic phonons (which are major participants in electron-phonon drag) by electrons in materials without inversion symmetry~\cite{kaasbjerg2013acoustic}. AlN is chosen also because of its hexagonal crystal structure with similar in-plane lattice parameters as $\rm MoS_2$. Surprisingly, we find that, at room temperature, the electron drag effect contributes up to 17\% increase in the lattice thermal conductivity of $\rm MoS_2$ at carrier concentrations between $\rm 10^{12}~cm^{-2}$ and $\rm 10^{14}~cm^{-2}$ by reducing phonon-electron scatterings through momentum circulation between electrons and phonons, while it has little impact on the thermal conductivity of AlN (smaller than 1\%). The electron drag effect in phosphorene is weaker but still increases its lattice thermal conductivity by nearly 4\%. Through detailed analysis of mode-resolved electron-phonon interaction in all three materials, we attribute the observed difference in the electron drag effect in 3D and 2D semiconductors to the larger phase space for phonon-electron scatterings and the higher contribution to the total thermal conductivity of drag-active long-wavelength acoustic phonons, both of which are intrinsic features in 2D semiconductors. The difference between MoS$_2$ and phosphorene is further determined to be due to the strong piezoelectric scattering in MoS$_2$. 
Our work demonstrates the significant contribution of the electron drag effect to thermal transport in 2D semiconductors and provides fundamental understanding of coupled electron-phonon transport in materials with varied dimensionalities.








\section{Results and Discussions}

\begin{figure}[!htb]
\includegraphics[width=0.8\textwidth]{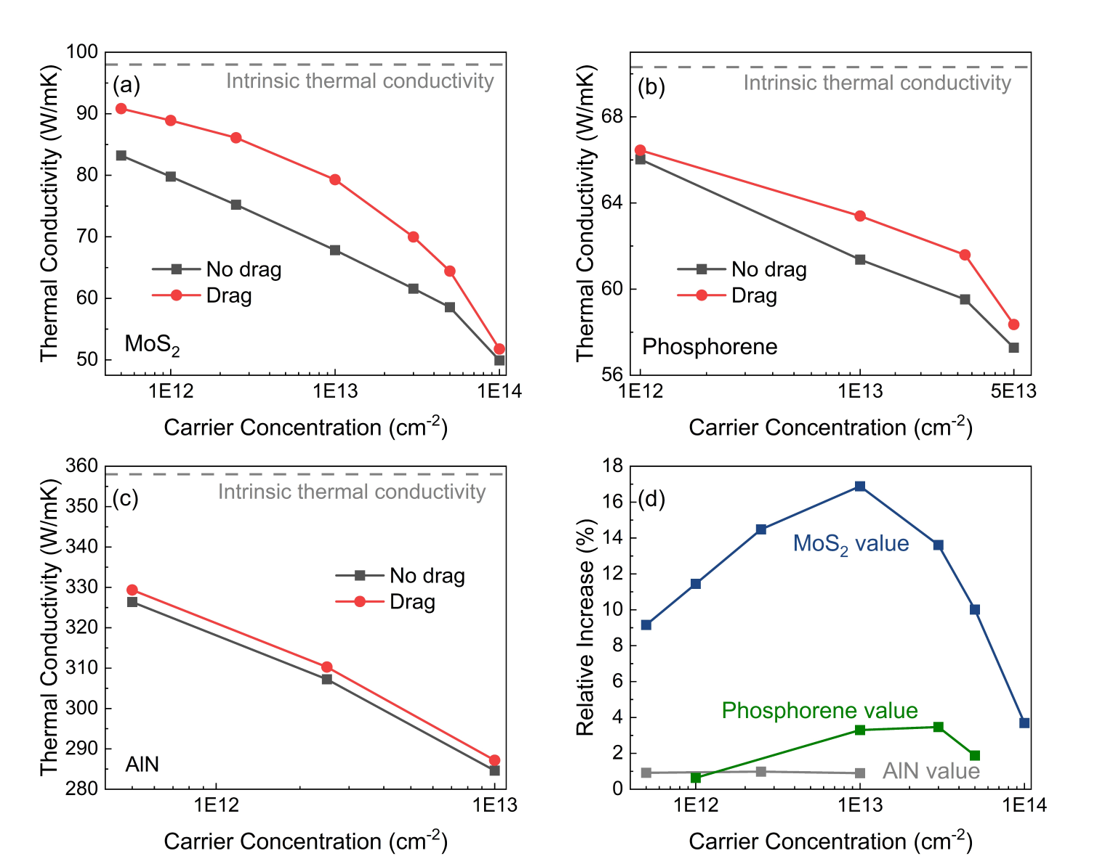}
\caption{\textbf{Impact of the electron drag effect on thermal conductivity.} Room temperature thermal conductivity of (a) n-type monolayer $\rm MoS_2$, (b) n-type monolayer phosphorene, and (c) n-type AlN as a function of carrier concentration with (red solid line) and without (black solid line) the electron drag effect. The gray dashed line denotes the calculated thermal conductivity of the pristine crystal without considering phonon-electron scatterings. (d) The relative thermal conductivity increase due to the electron drag effect. The blue line shows the relative increase in $\rm MoS_2$, the green line shows the relative increase in phosphorene and the gray line shows the relative increase in AlN. The electron drag effect is more predominant in 2D materials.} 
\label{fig:fig1}
\end{figure}

Details of the electron-phonon drag theory and computational methods are provided in the Supporting Information. The calculated room temperature thermal conductivities of monolayer n-type $\rm MoS_2$, monolayer n-type phosphorene, and n-type AlN as a function of carrier concentrations are shown in Fig.~\ref{fig:fig1}.
The gray dashed line shows the intrinsic thermal conductivity without any phonon-electron scatterings, and the value is averaged over the anisotropic directions in phosphorene and AlN. 
The calculated thermal conductivity values agree well with other calculations \cite{gu2016layer,jain2015strongly,cheng2020experimental}.
The thermal conductivity decreases in all materials with increasing carrier concentration due to the increased phonon-electron scatterings.
If the electron drag is considered, the thermal conductivity of monolayer $\rm MoS_2$ shows a significant increase compared with the case where the electron drag effect is neglected, as shown in Fig.~\ref{fig:fig1}(a). 
The electron drag also has an evident impact on the thermal conductivity of phosphorene, as shown in Fig.~\ref{fig:fig1}(b).
To compare with 3D semiconductors, the electron drag effect on thermal conductivity in AlN is shown in Fig.~\ref{fig:fig1}(c), where the 3D carrier concentration is converted to an equivalent 2D value by multiplying the out-of-plane lattice parameter.
The relative increase in thermal conductivity due to the electron drag effect in all three materials is summarized in Fig.~\ref{fig:fig1}(d).
At $\rm n = 10^{13}~cm^{-2}$, the thermal conductivity of $\rm MoS_2$ is increased by 17\% due to the electron drag, and the thermal conductivity is increased by 3.6\% in phosphorene. 
In both materials, the relative change in thermal conductivity first increases with an increasing carrier concentration, which can be explained by the fact that an increasing carrier concentration enables more electron-phonon scattering channels and more momentum circulation between electrons and phonons.
Since the calculation also includes the electron-charged-impurity scattering, the percentage contribution of the electron drag decreases at higher carrier concentration due to the increased dissipation of electron momentum by charged impurities.
Although the electron drag effect on thermal conductivity is evident in both 2D semiconductors, it has little influence on AlN.
As shown in Fig.~\ref{fig:fig1}(d), the relative thermal conductivity increase due to the electron drag is less than 1\% in the entire carrier concentration range.
This negligible electron drag effect in AlN agrees well with other calculations of 3D materials, such as SiC \cite{protik2020electronSiC} and GaAs \cite{protik2020coupledGaAs} (when the electron-charged-impurity scattering is also considered), where the weak contribution of electron drag is attributed to the small fraction of phonons that are drag active compared with the wide spectrum of phonons that contribute to the thermal conductivity. 

\begin{figure}[!htb]
\includegraphics[width=1\textwidth]{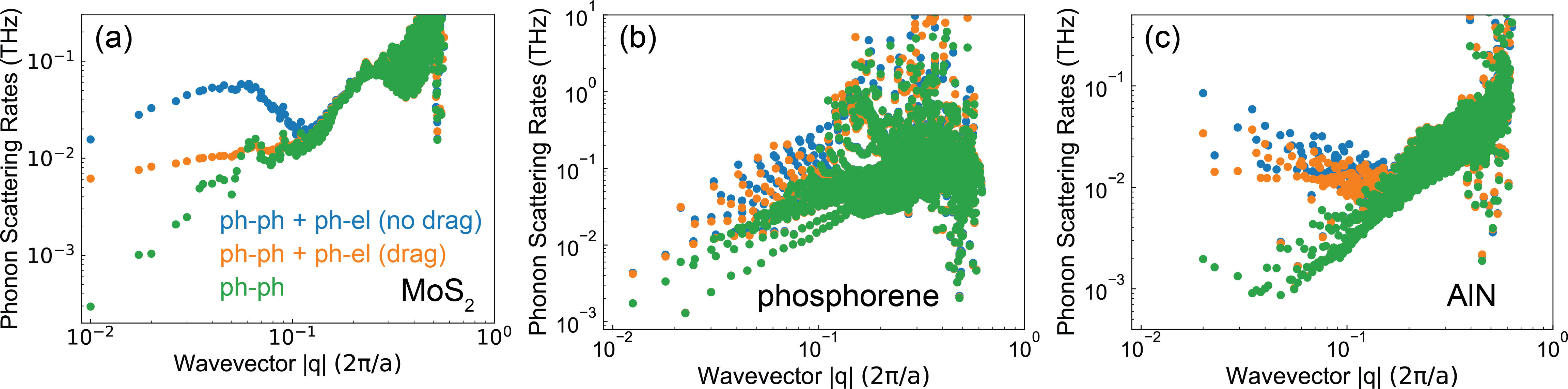}
\caption{\textbf{Impact of electron drag on the scattering rates of selected phonon modes.} Phonon scattering rates as a function of wavevector $|\mathbf{q}|$ for (a) the LA mode in $\rm MoS_2$, (b) the LA mode in phosphorene, and (c) the lowest TA mode in AlN. The green dots denote the phonon-phonon scattering rates, the blue dots denote the phonon-phonon and phonon-electron scattering rates without considering the electron drag effect, and the orange dots represent the total phonon-phonon and phonon-electron scattering rates when the electron drag is considered. Electron drag decreases phonon-electron scattering in all three materials.} 
\label{fig:fig2}
\end{figure}

To have a better understanding of the reason why electron drag has a much stronger impact on 2D semiconductors, the phonon scattering rates of different branches are analyzed as a function of the phonon wavevector. 
For all three materials that we investigated, the contribution of the electron drag effect decomposed into each phonon branch is shown in Fig.~S3.
It is found that, in $\rm MoS_2$, the LA mode contributes to more than 85\% of the total thermal conductivity increase due to the electron drag over all carrier concentrations. In phosphorene, both the LA mode and the lowest optical phonon mode significantly contribute to the increase in the thermal conductivity due to the electron drag. Whereas in AlN, the first TA mode dominates the electron drag contribution.
Therefore, the phonon scattering rates of these modes at $ \rm n = 10^{13}~cm^{-2}$ are shown in Fig.~\ref{fig:fig2} to better illustrate the impact of the electron drag effect microscopically.
The phonon scattering rates of other acoustic phonon modes are provided in Fig.~S4.
The green dots represent the phonon-phonon scattering rates, and the blue dots show the total scattering rates without considering the electron drag, which include both phonon-phonon scatterings and phonon-electron scatterings.
The difference between the phonon-phonon scattering rate and the total phonon scattering rate indicates the phonon-electron scattering rate.  
The orange dots are the total phonon scattering rates when the electron drag effect is considered, and it is found that the phonon-electron scattering rate of the LA mode in $\rm MoS_2$ is significantly decreased due to the electron drag, while it has little change in AlN.
For phosphorene, the phonon-electron scattering is quite weak even when the electron drag effect is not considered, which is also shown in Fig.~\ref{fig:fig1}(b), where the thermal conductivity of phosphorene decreases much slower with an increasing carrier concentration compared to $\rm MoS_2$ and AlN. 
Therefore, the electron drag effect on the phonon-electron scattering rate in phosphorene, which originates from momentum circulation through electron-phonon interactions, is less significant than that in $\rm MoS_2$.
 

\begin{figure}[!htb]
\includegraphics[width=1\textwidth]{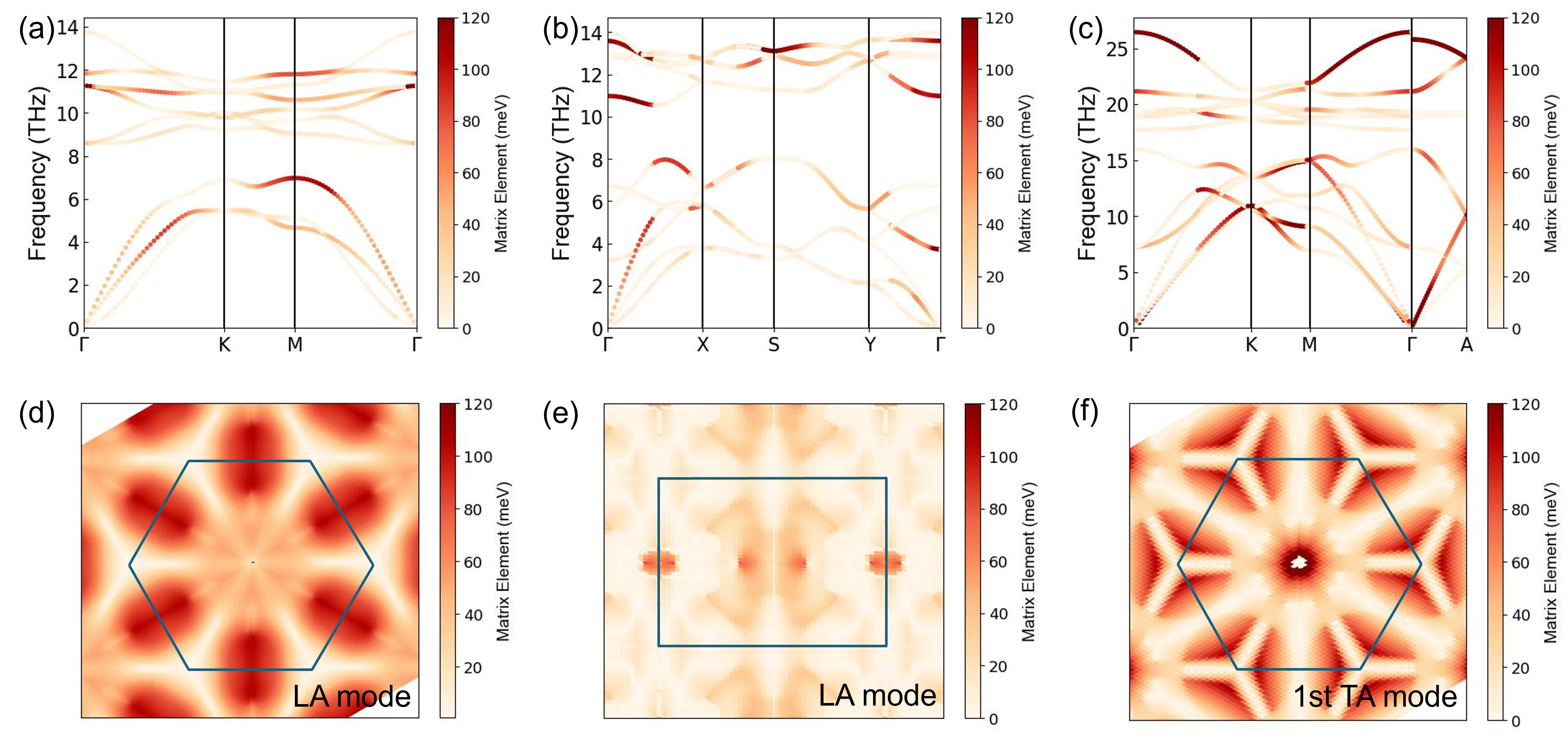}
\caption{\textbf{Electron-phonon scattering matrix elements associated with selected phonon modes.} The first row shows the phonon dispersions of (a) $\rm MoS_2$, (b) phosphorene, and (c) AlN. The color indicates the magnitude of electron-phonon matrix elements. The second row shows the electron-phonon matrix elements in reciprocal space for (d)  the LA mode in $\rm MoS_2$, (e) the LA mode in phosphorene, and (f) the first TA mode in AlN. The blue lines delineate the boundary of the first Brillouin zone. Phosphorene has the smallest matrix elements, while the matrix elements in $\rm MoS_2$ and AlN are larger due to piezoelectric scattering.} 
\label{fig:fig3}

\end{figure}

To gain more insight into the microscopic mechanism of the electron drag effect on thermal conductivity that originates from electron-phonon interactions, we start with the equation that provides the phonon-electron scattering rate \cite{ziman2001electrons}:
\begin{equation}
    \begin{aligned}
        \frac{1}{\tau_{s\mathbf{q}}} = \frac{2\pi}{\hbar}\frac{2}{1+n_{s\mathbf{q}}^0} \sum_{mn\mathbf{k}}\lvert g^{smn}_{\mathbf{k}\mathbf{q}}\rvert^2 f^0_{m\mathbf{k}}(1-f^0_{n\mathbf{k+q}})\delta(\varepsilon_{n\mathbf{k+q}}-\varepsilon_{m\mathbf{k}}-\hbar \omega_{s\mathbf{q}}),
    \end{aligned}
\label{eq:1}
\end{equation}
where $\lvert g^{smn}_{\mathbf{k}\mathbf{q}}\rvert$ is the electron-phonon matrix element describing the scattering of an electron from the initial state $m\mathbf{k}$ to the final state $n\mathbf{k+q}$ with a phonon of branch $s$ and wavevector $\mathbf{q}$, $f^0$ is the Fermi-Dirac distribution of the electron, $n_0$ is the Bose-Einstein distribution of the phonon, and the factor 2 in $\dfrac{2}{1+n_{s\mathbf{q}}^0}$ accounts for the spin degeneracy.
According to Eq.~\ref{eq:1}, the interaction between phonons and electrons is determined by both electron-phonon matrix elements, which signal the strength of the coupling between one phonon and two electrons, and the number of electron-phonon pairs that satisfy energy and momentum conservation.
The matrix elements associated with each phonon mode calculated with the initial electronic state wavevector $\mathbf{k}$ located at the conduction band minimum are shown in Fig.~\ref{fig:fig3}(a-c) and the matrix elements associated with the LA mode in MoS$_2$ and phosphorene, and the first TA mode in AlN are further plotted in the Brillouin zone in Fig.~\ref{fig:fig3}(d-f).
The matrix elements of the three materials are plotted using the same scale. For the electron-phonon drag effect, the most important phonon modes are the long-wavelength acoustic modes located near the $\Gamma$ point. It is seen that the matrix elements of acoustic modes near $\Gamma$ in $\rm MoS_2$ are larger than those in phosphorene. 
The dominant mechanism of interactions between electrons and acoustic phonons in phosphorene originates from the deformation potential \cite{bardeen1950deformation}, which is the change of the band edge energies due to lattice perturbations from phonons. 
In contrast, for $\rm MoS_2$, which is a polar material without inversion symmetry, in addition to the deformation potential scattering, the piezoelectric scattering of electrons by acoustic phonons also plays a key role \cite{meijer1953note,kaasbjerg2013acoustic}. 
Therefore, the larger electron-phonon matrix elements due to piezoelectric scattering enable a stronger contribution to the electron drag in $\rm MoS_2$ compared with phosphorene.
Piezoelectric scattering is also significant in AlN, which is polar and without inversion symmetry. In fact, piezoelectric scattering generally leads to a large and finite matrix element for acoustic phonons as $\mathbf{q}$ approaches 0 in 3D polar semiconductors without inversion symmetry~\cite{jhalani2020piezoelectric}, which is clearly shown in Fig.~\ref{fig:fig3}(c) and (f). Although the matrix elements are very large in AlN near $\Gamma$, they also decay quickly as $\mathbf{q}$ increases, and thus the strong electron-phonon interaction is limited to long-wavelength acoustic phonons near the zone center. 

\begin{figure}[!htb]
\includegraphics[width=1\textwidth]{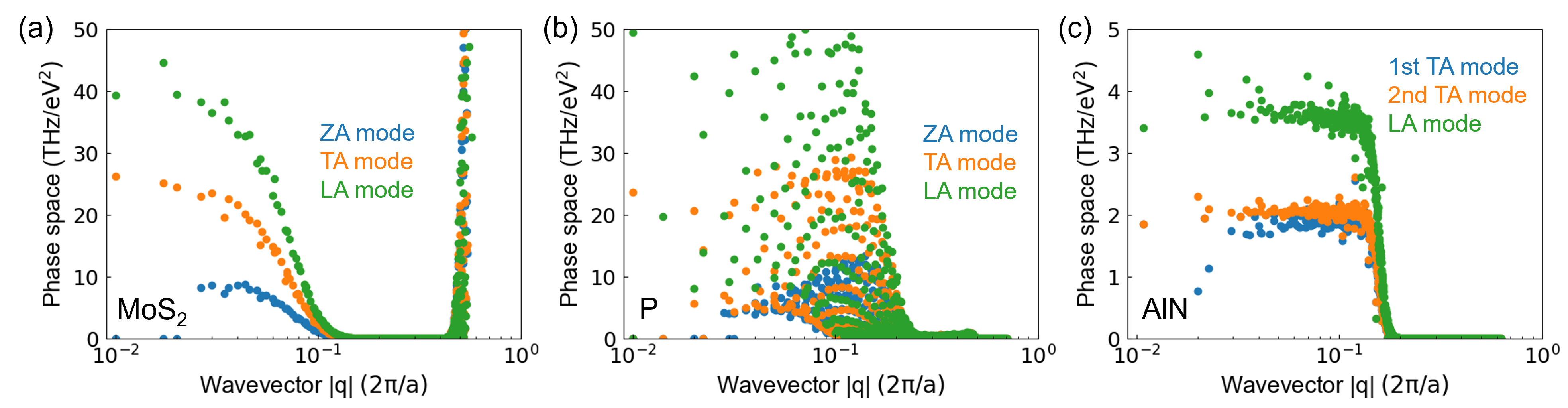}
\caption{\textbf{Electron-phonon scattering phase space of acoustic phonon modes}. Electron-phonon scattering phase space as a function of phonon wavevector in (a) $\rm MoS_2$, (b) phosphorene, and (c) AlN. The scattering phase space in 2D MoS$_2$ and phosphorene is one order of magnitude higher than that in 3D AlN.} 
\label{fig:fig4}
\end{figure}

In addition to the matrix elements, the number of electron and phonon states that satisfy the energy and momentum conservation conditions also determines the strength of electron-phonon interaction and can be quantified by the phonon-electron scattering phase space $P_{s\mathbf{q}}$, in which the electron-phonon matrix element in Eq.~\ref{eq:1} is set to unity:
\begin{equation}
    \begin{aligned}
        P_{s\mathbf{q}} = \frac{2\pi}{\hbar}\frac{2}{1+n_{s\mathbf{q}}^0} \sum_{mn\mathbf{k}} f^0_{m\mathbf{k}}(1-f^0_{n\mathbf{k+q}})\delta(\varepsilon_{n\mathbf{k+q}}-\varepsilon_{m\mathbf{k}}-\hbar \omega_{s\mathbf{q}}).
    \end{aligned}
\label{eq:2}
\end{equation}
The calculated electron-phonon phase space of acoustic modes is shown in Fig.~\ref{fig:fig4} for $\rm MoS_2$, phosphorene, and AlN at $\rm n = 10^{13}~cm^{-2}$, respectively.
The phase space of 2D MoS$_2$ and phosphorene is one order of magnitude higher than that of AlN, and the large phase space at long wavevectors in $\rm MoS_2$ is due to intervalley scatterings. The large electron-phonon scattering phase space in 2D semiconductors has been discussed in the context of understanding the generally low intrinsic electron mobility in 2D semiconductors and was attributed to the large electronic density of states (DOS) and the lower phonon energy in 2D~\cite{cheng2020two}.
The large phase space offers greater opportunities for momentum to circulate back to phonons from electrons, which was previously lost due to scatterings by electrons, leading to less dissipation of phonon momentum and resulting in a noticeable electron drag effect on thermal conductivity in 2D semiconductors. 
In contrast, although AlN has large electron-phonon matrix elements, as shown in Fig.~\ref{fig:fig3}(c) and (f), due to the much smaller phase space compared with MoS$_2$ and phosphorene, as shown in Fig.~\ref{fig:fig4}, the electron drag effect is weak.

\begin{figure}[!htb]
\includegraphics[width=0.75\textwidth]{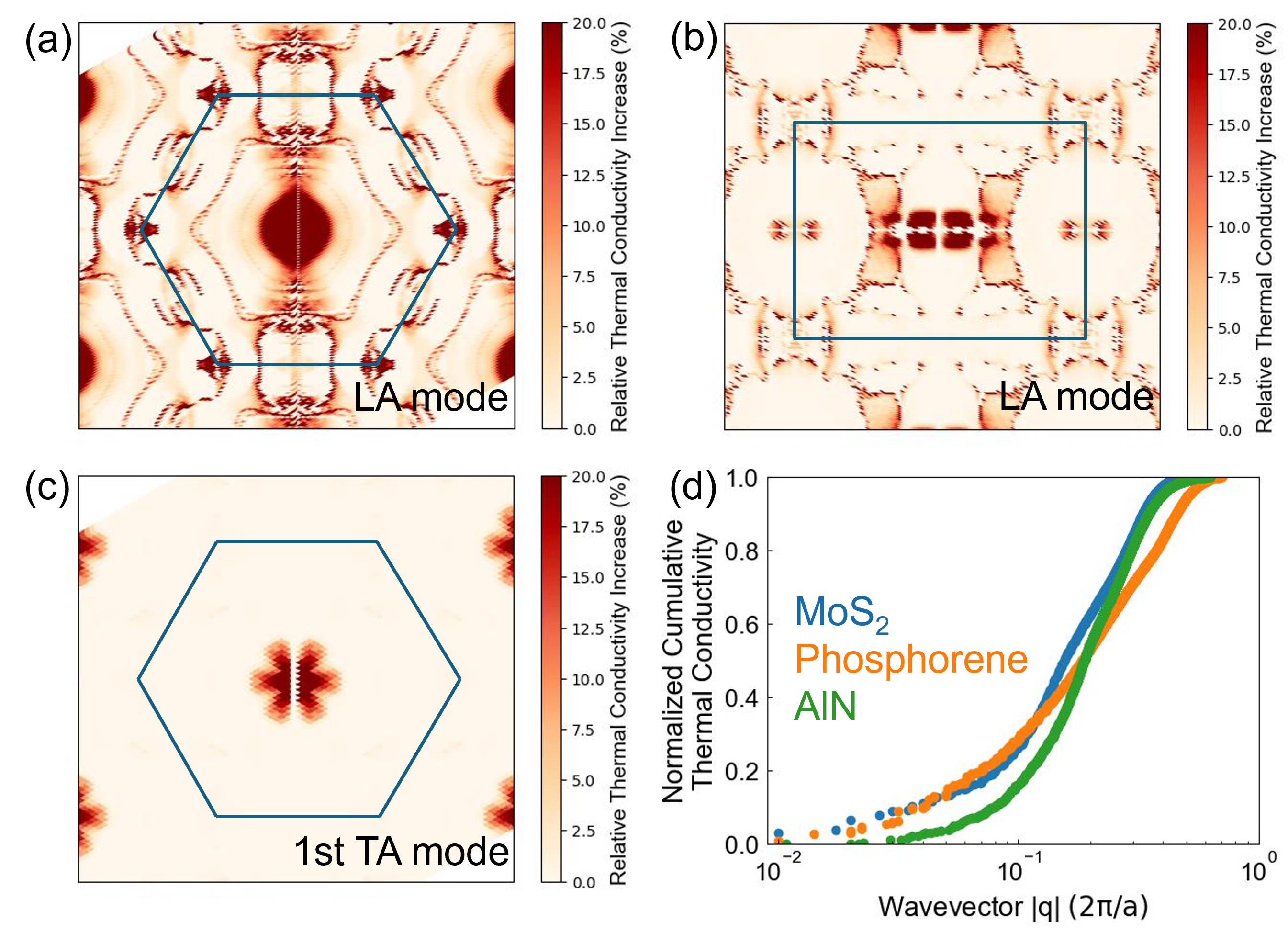}
\caption{\textbf{Impact of electron drag on mode-resolved thermal conductivity of selected phonon modes.} Relative thermal conductivity change due to electron drag of (a) the LA mode in $\rm MoS_2$, (b) the LA mode in phosphorene, and (c) the first TA mode in AlN. The blue lines delineate the boundary of the first Brillouin zone. (d) Normalized accumulated contribution to the total thermal conductivity as a function of phonon wavevector. Drag-active long-wavelength phonons with wavevectors smaller than 0.1\,$\frac{2\pi}{a}$ contribute more to the total thermal conductivity in 2D materials than in 3D materials.} 
\label{fig:fig5}
\end{figure}

Figure~\ref{fig:fig5}(a-c) shows the relative change in thermal conductivity due to electron drag of selected phonon modes in MoS$_2$, phosphorene, and AlN. In all three materials, drag-active phonons are mainly concentrated near the $\Gamma$ point ($q<0.1 \frac{2\pi}{a}$, where $a$ is the lattice parameter) as limited by the scattering phase space shown in Fig.~\ref{fig:fig4}. These low-frequency, long-wavelength phonons, however, contribute differently to the total thermal conductivity in 2D and 3D due to the fact that the phonon density of states scales with $\omega$ in 2D and $\omega^2$ in 3D. This effect is shown in Fig.~\ref{fig:fig5}(d), where normalized accumulated contribution to the total thermal conductivity as a function of phonon wavevector is shown. It is clearly seen that drag-active long-wavelength phonons contribute significantly more to the total thermal conductivity in 2D MoS$_2$ and phosphorene than in 3D AlN. In addition to room temperature, the electron drag effect on thermal conductivity at lower temperatures is shown in Fig.~S5. 
At lower temperatures, the electron drag effect in 2D semiconductors becomes more evident at lower carrier concentrations, which is a consequence of more dominant phonon-electron scattering over phonon-phonon scattering, enabling more momentum flowing back to phonons, while at higher concentrations, the electron-charged-impurity scatterings limit this process. In contrast, the electron drag effect in 3D semiconductors has little change at low temperatures, which can be explained by the small electron-phonon scattering phase space limiting the momentum circulation.

In summary, we have systematically studied the electron drag effect on thermal conductivity in 2D semiconductors by solving the fully coupled electron-phonon BTEs. 
We find that the electron drag effect can significantly increase the thermal conductivity by reducing the phonon-electron scattering in 2D semiconductors, while this effect has negligible influence in 3D. We attribute this distinction to the larger electron-phonon scattering phase space and the higher contribution to the total thermal conductivity of drag-active long-wavelength acoustic phonons in 2D. We note here that both features are generally present in 2D semiconductors~\cite{cheng2020two}, suggesting that our conclusions can be generalized to other 2D semiconductors. We also find the electron drag effect is particularly strong in 2D polar semiconductors without an inversion symmetry, such as the family of transition-metal dichalcogenides, where piezoelectric scattering plays an important role. Our work advances the fundamental understanding of coupled electron-phonon transport in 2D materials and highlights the importance of including the electron drag effect in thermal transport studies of 2D materials. 

\begin{acknowledgments}
We thank Dr. Nakib H. Protik for assistance with the Elphbolt code and Yubi Chen for helpful discussions. This work is based on research supported by the U.S. Air Force Office of Scientific Research under award number FA9550-22-1-0468. Y.Q. also acknowledges support from the NSF Quantum Foundry via the Q-AMASE-i program under award number DMR-1906325 at the University of California, Santa Barbara (UCSB). This work used Stampede2 at Texas Advanced Computing Center (TACC) through allocation MAT200011 from the Advanced Cyberinfrastructure Coordination Ecosystem: Services \& Support (ACCESS) program, which is supported by National Science Foundation grants 2138259, 2138286, 2138307, 2137603, and 2138296. Use was also made of computational facilities purchased with funds from the National Science Foundation (award number CNS-1725797) and administered by the Center for Scientific Computing (CSC) at University of California, Santa Barbara (UCSB). The CSC is supported by the California NanoSystems Institute and the Materials Research Science and Engineering Center (MRSEC; NSF DMR-2308708) at UCSB. 
\end{acknowledgments}

\bibliography{references.bib}

\end{document}



\title{Supporting Information: Electron Drag Effect on Thermal Conductivity in Two-dimensional Semiconductors} 

\author{Yujie Quan}
\affiliation{Department of Mechanical Engineering, University of California, Santa Barbara, CA 93106, USA}

\author{Bolin Liao}
\email{bliao@ucsb.edu} \affiliation{Department of Mechanical Engineering, University of California, Santa Barbara, CA 93106, USA}

\maketitle





\section{Computational Methods}




The detailed theory of the coupled electron-phonon BTEs was discussed in Ref.~\cite{quan2023significant}.
In short, the electron drag effect originates from the collision term in phonon BTE due to electron-phonon interactions:
\begin{equation}
    \begin{aligned}
        \left(\dfrac{\partial n_{s\mathbf{q}}}{\partial t}\right)_{e-p h} \simeq &\sum_{m\mathbf{k}, n\mathbf{k}^{\prime}}\left[G_{m\mathbf{k}}\left(n\mathbf{k}^{\prime} , s\mathbf{q} \right) \cdot \Delta f_{m\mathbf{k} }+G_{n\mathbf{k}^{\prime}}(m\mathbf{k} , s\mathbf{q}) \cdot \Delta f_{n\mathbf{k}^{\prime} }\right]-\\ &\left[\sum_{m\mathbf{k}, n\mathbf{k}^{\prime}} G_{s\mathbf{q}}\left(m\mathbf{k}, n\mathbf{k}^{\prime} \right)\right] \cdot \Delta n_{s\mathbf{q}},
    \end{aligned}
\label{eq:0}
\end{equation}
where coefficients $G$ only depend on the equilibrium electron distribution $f^0$ and phonon distribution $n^0$. In typical ``decoupled'' phonon transport calculations~\cite{liao2015significant}, only the second term is evaluated, where the electron-phonon scattering rate is computed based on $f^0$ and $n^0$ and included in $G_{s\mathbf{q}}$. 
The first term in Eq.~\ref{eq:0}, related to the nonequilibrium electron distribution $\Delta f$, corresponds to the electron drag effect, indicating nonequilibrium electrons have impacts on phonon distributions, and is fully evaluated in the coupled BTE approach~\cite{protik2020coupledGaAs}.

DFT calculations are accomplished via the Quantum ESPRESSO package \cite{giannozzi2009quantumQE} with the scalar-relativistic optimized norm-conserving Vanderbilt (ONCV) pseudopotentials \cite{hamann2013optimizedONCV} within the local density approximation (LDA) \cite{perdew1992accurateLDA}. 
The kinetic energy cutoff for wave functions was set to 80 Ry for all of our calculations. 
The total electron energy convergence threshold for self-consistency was set to $\rm 1 \times 10^{-10}$ Ry. 
A mesh grid of $\rm 12\times 12\times 12$ in the first Brillouin zone was adopted for the calculations of AlN, and a mesh grid of $18\times 18 \times 1$ and $12\times 12 \times 1$ was adopted for $\rm MoS_2$ and phosphorene, respectively. 
The lattice is fully relaxed with a force threshold of $\rm 10^{-4}~eV/\textup{\AA}$, with lattice parameters $\rm a = 3.12~\textup{\AA}$, $\rm c = 5.0 ~\textup{\AA}$ in AlN, $\rm a = 3.18 ~\textup{\AA}$ in $\rm MoS_2$, and $\rm a = 3.27 ~\textup{\AA}$, $\rm b = 4.36 ~\textup{\AA}$ in phosphorene, all of which were in excellent agreement with the experimental values \cite{schulz1977crystal,vancso2016intrinsic,jain2015strongly}.
To avoid interactions between layers, a vacuum layer of $\rm 20~ \textup{\AA}$ was introduced along the axis perpendicular to the $\rm MoS_2$ and phosphorene monolayer. 
The phonon dispersions and second-order IFCs were calculated using DFPT \cite{baroni2001phononsDFPT} as implemented in Quantum ESPRESSO. 
A $6 \times 6 \times 6$ q-point grid was used for the phonon calculation of AlN, and a $6 \times 6 \times 1$ q-point grid was used for $\rm MoS_2$ and phosphorene.
The third-order IFCs were computed using a $3 \times 3 \times 3$ supercell for AlN, and a $6 \times 6 \times 1$ supercell for $\rm MoS_2$ and phosphorene with the finite displacement method \cite{li2014shengbte}, taking up to the fifth nearest neighbors into consideration.
The electron-phonon matrix elements used for future coupled BTEs calculations were calculated using the EPW package \cite{ponce2016epw}, with a coarse $12 \times 12 \times 12$ k-point grid and a $6 \times 6 \times 6$ q-point grid for AlN, a $18 \times 18 \times 1$ k-point grid and a $6 \times 6 \times 1$ q-point grid for $\rm MoS_2$, and a $12 \times 12 \times 1$ k-point grid and a $6 \times 6 \times 1$ q-point grid for phosphorene.
The Wannier-interpolated electronic band structures and the ones calculated from first-principles are compared in Fig.~S1 with good agreement.
The output of the electron-phonon matrix elements in Wannier space was used as an input for transport calculations using Elphbolt \cite{protik2022elphbolt}, where the real-space Wannier representation quantities were transformed to the Bloch representations on a fine $50 \times 50 \times 50$ q-point grid and a $150 \times 150 \times 150$ k-point grid for AlN, and a $100 \times 100 \times 1$ q-point grid and a $300 \times 300 \times 1$ k-point grid for $\rm MoS_2$ and phosphorene.
The convergence of the calculated results with respect to the q-point grid density is shown in Fig.~S2.
The transport calculations also considered the phonon-isotope scatterings using the Tamura model \cite{tamura1983isotope}, and electron-charged-impurity scattering, in which the Brooks-Herring model \cite{brooks1955theory} was adopted, treating the impurity as a static, Yukawa-type scattering center. 

\clearpage

\section{Supplementary Figures}

\begin{figure}[!htb]
\includegraphics[width=\textwidth]{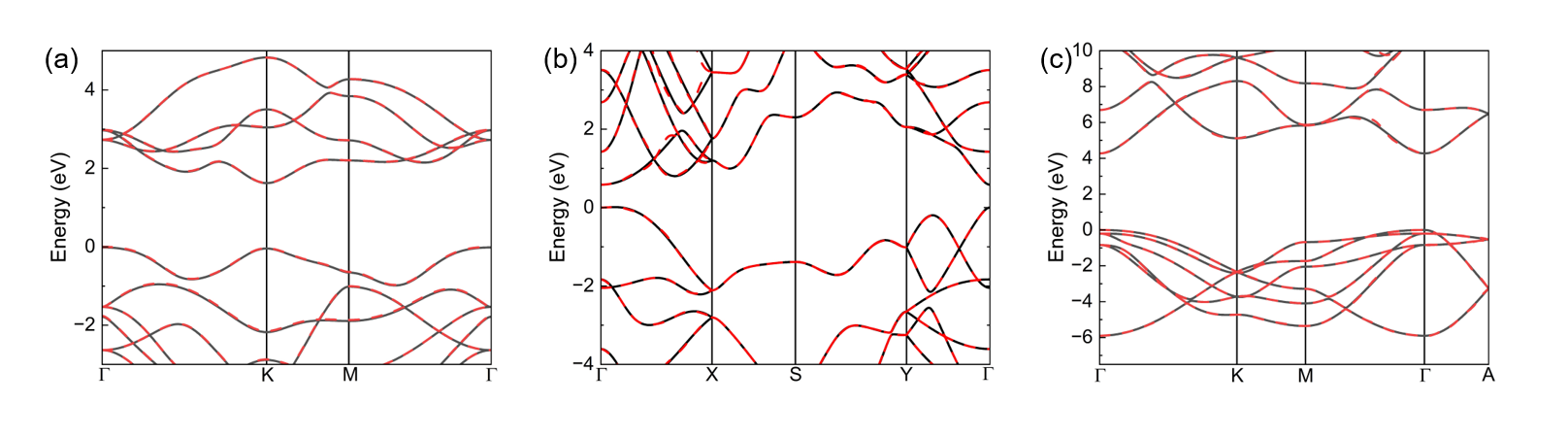}
\caption{Electronic band structures of (a) $\rm MoS_2$, (b) phosphorene, and (c) AlN. The black solid lines correspond to the direct DFT calculation, and the red dashed lines show the Wannier-interpolated band structures.} 
\label{fig:figS1}
\end{figure}
\clearpage

\begin{figure}[!htb]
\includegraphics[width=0.6\textwidth]{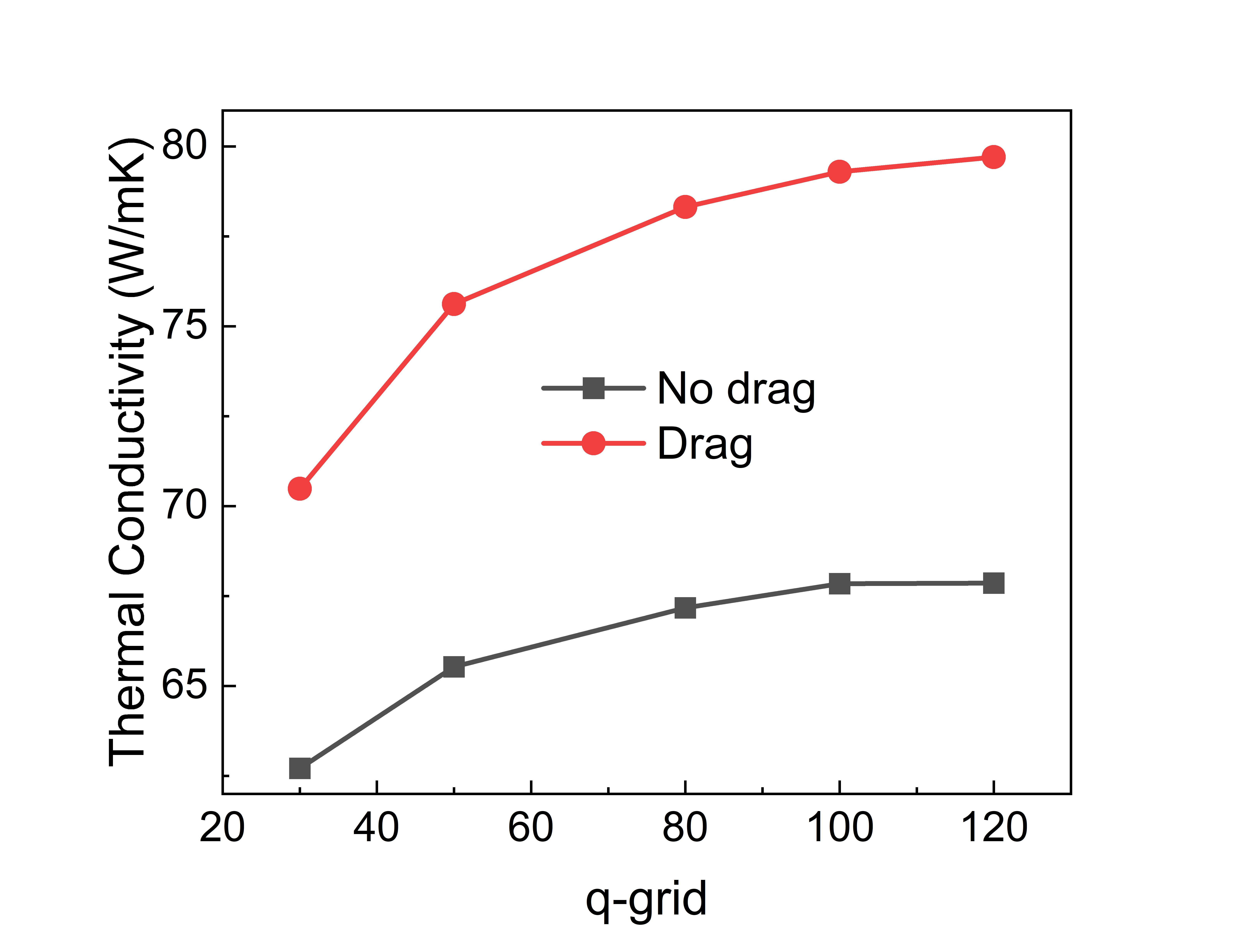}
\caption{Room-temperature thermal conductivity of $\rm MoS_2$ at $\rm n = 10^{13}~cm^{-2}$ calculated as a function of q-grid sampling density. The calculated thermal conductivity with and without the electron drag effect tends to converge at $100 \times 100 \times 1$ q-grid density.}
\end{figure}
\clearpage

\begin{figure}[!htb]
\includegraphics[width=\textwidth]{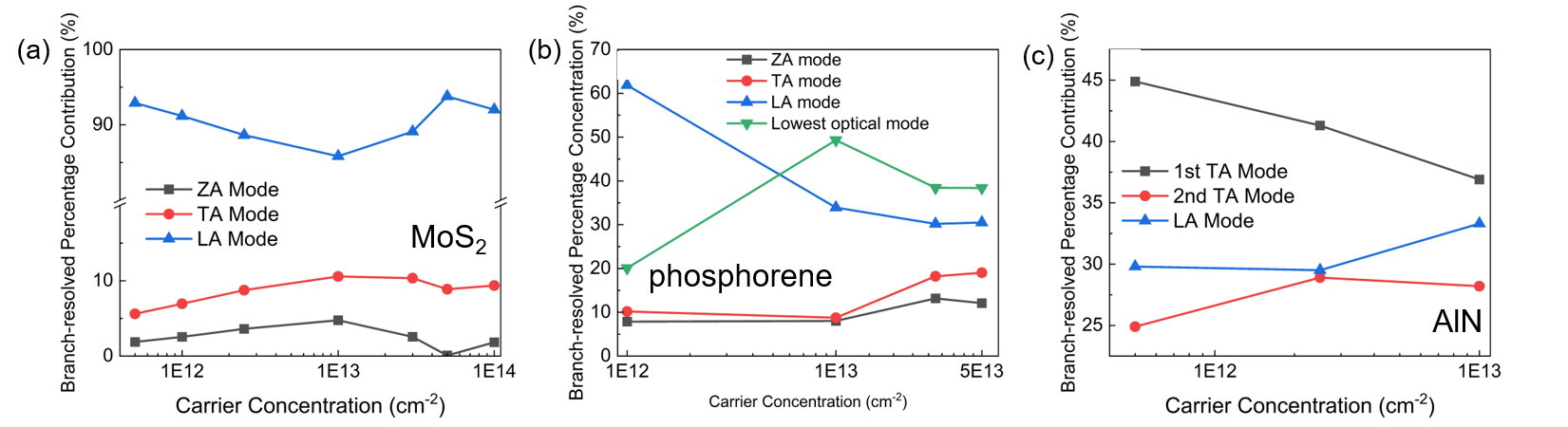}
\caption{Branch-resolved electron drag percentage contribution to total thermal conductivity increase in (a) $\rm MoS_2$, (b) phosphorene, and (c) AlN. The LA mode contributes the most in $\rm MoS_2$, the LA and the lowest optical mode contribute the most in phosphorene, and the lowest TA mode contributes the most in AlN.}
\end{figure}
\clearpage

\begin{figure}[!htb]
\includegraphics[width=0.75\textwidth]{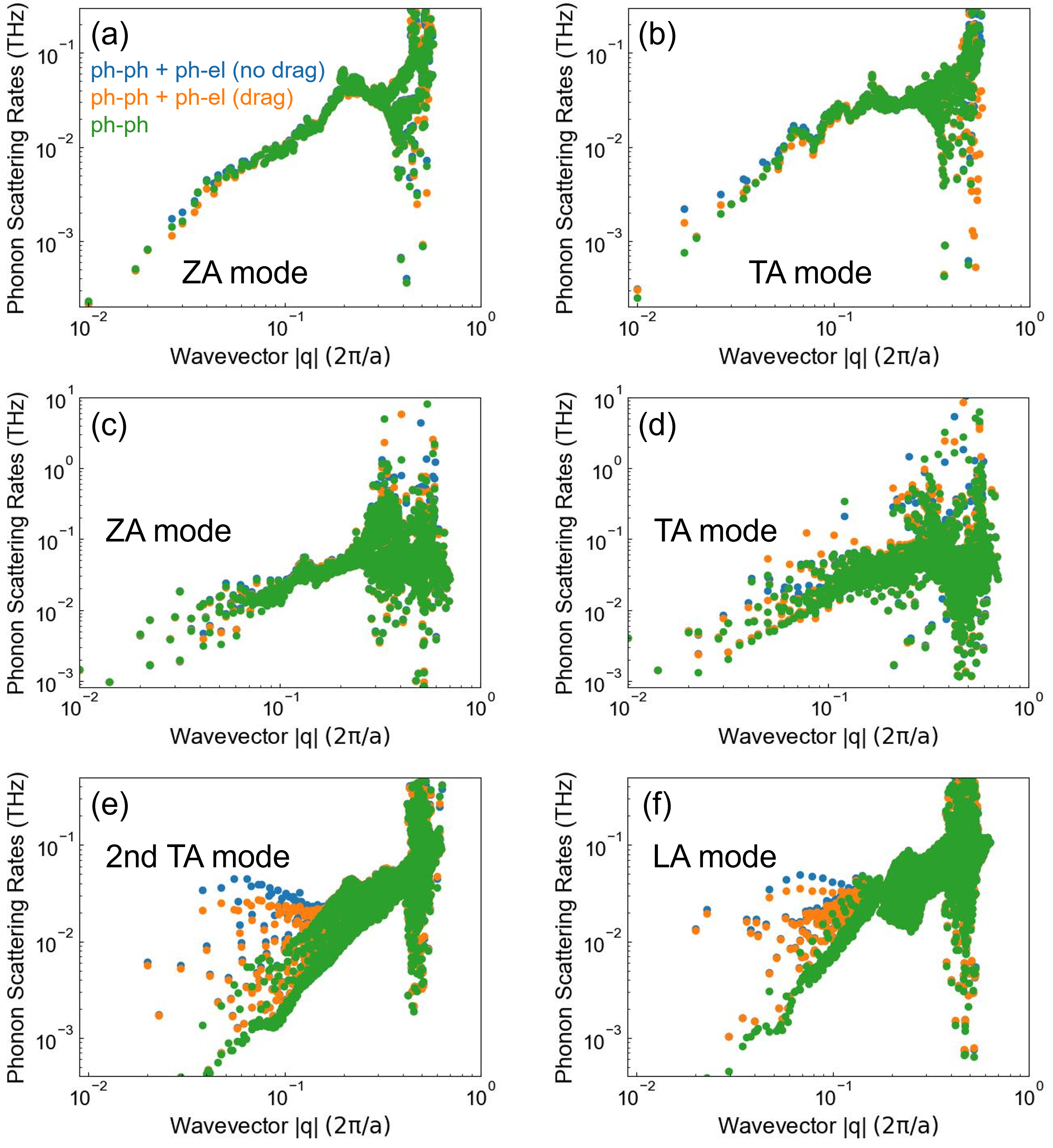}
\caption{Phonon scattering rates of other acoustic phonon modes that are not shown in the main text. The first, second, and third rows correspond to $\rm MoS_2$, phosphorene, and AlN, respectively. }
\end{figure}
\clearpage

\begin{figure}[!htb]
\includegraphics[width=0.6\textwidth]{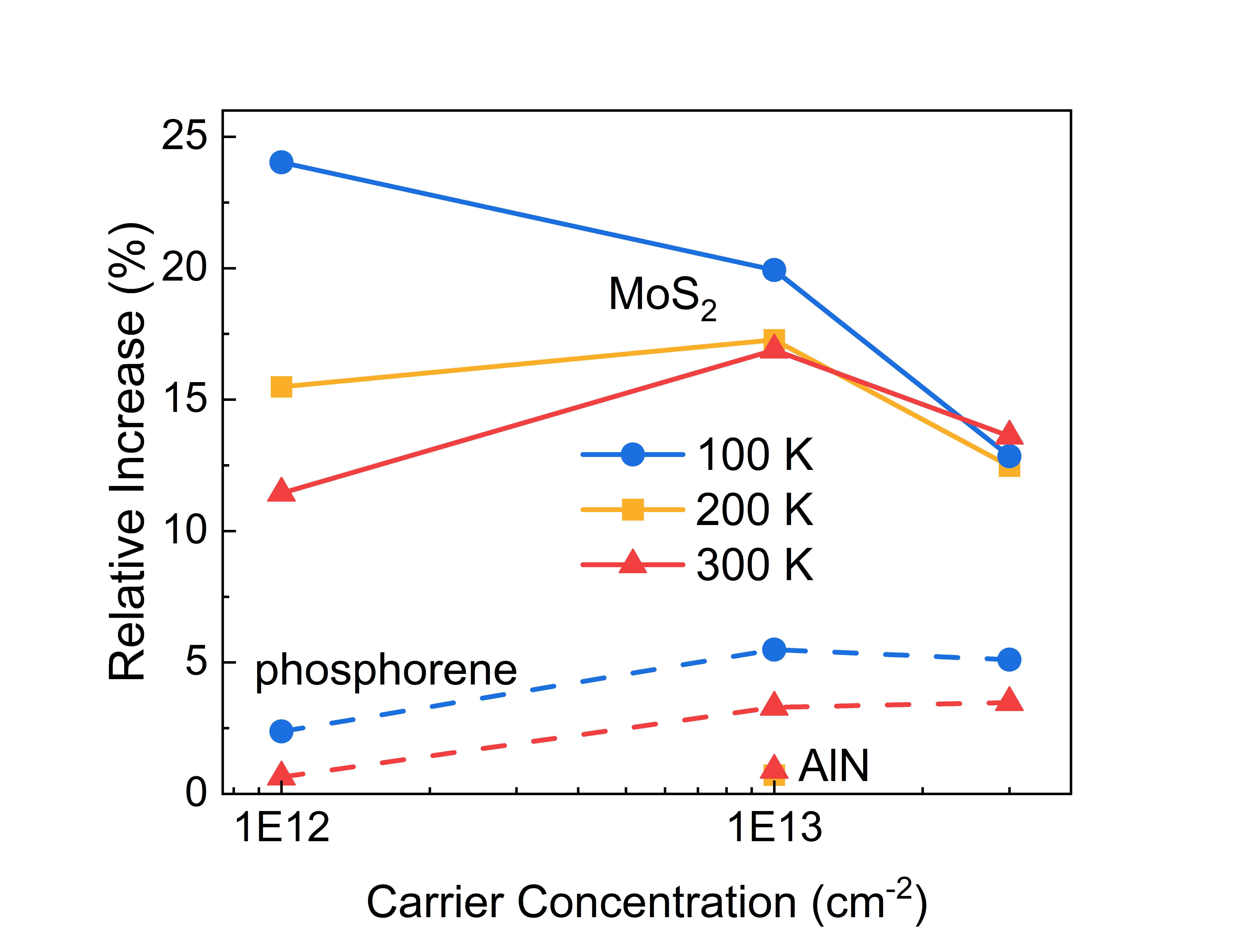}
\caption{Electron drag effect on thermal conductivity at different temperatures. The red line shows the results at 300 K, the yellow line corresponds to 200 K, and the blue line shows the results at 100 K. At low temperatures, the electron drag effect in 2D materials becomes more evident at lower carrier concentrations, while in 3D, it has little change.}
\end{figure}
\clearpage

\section*{References}
\bibliography{references.bib}%